\documentclass[12pt]{article}
\usepackage{graphics}
\begin{document}
\begin{center}
{\large{\bf Soliton in Gravitating Gas. Hoag's Object.}}

\vspace*{10mm}

{\bf G.P. Pronko}\\
{\it IHEP, Protvino, Moscow reg., Russia}

\begin{abstract}
We explore the possibility of creating of solitons in gravitating
gas. It is shown that the virial arguments does not put an obstacle
for the existence of localized static solutions. The simplest
toroidal soliton of gravitating gas could be the explanation of the
peculiar galaxy named Hoag's object.
\end{abstract} 
\end{center}

\section{Introduction}

In the recent paper  \cite{P.A.} we have considered
the Hamiltonian formalism for fluid and gas based on the
Lagrangian description. It was pointed out in this paper that apart
from
other advantages, the Lagrangian description, which uses the
trajectories
of the particles of fluid (gas) as the dynamical variables, is the
most
convenient  for the introduction of interaction. In particular it was
demonstrated in \cite{P.A.} how the introduction of the
electromagnetic
interaction of particles which constitute the fluid, provide us with
the
theory of plasma. In the present paper we are going to consider in
analogous way the theory of gas of particles which interact with each
other through gravitational Newton potential. The system of such
particles
could be considered as a model for the motion of stars in a galaxy
when
the gravitation interaction prevails all other interaction. The total
number of stars in typical galaxy is of the order of
$10^{13}-10^{14}$, so
it may be reasonable to consider this collection of "particles" as a
gas.

The simplest model, which is usually used for numerical simulation of
N-body model of galaxy is described by the Hamiltonian
\begin{eqnarray}
H=\sum_{i=1}^{N}\frac{\vec
p{_i}^2}{2m_i}-\gamma\sum_{i\not=j}^{N}\frac{1}{|\vec x_i-\vec
x_j|} , \nonumber
\end{eqnarray}
where $\vec x_i, \vec p_i$ are canonical coordinates of particles
(stars)
with masses $m_i$. The model we are going to consider is based on the
dynamics described by $H$ with the assumption $m_i=m$, when
$N\rightarrow
\infty$. For this limit there appears a natural desire to consider  a
continuos distribution of the particles as it is done in the theory
of fluid or
gas \cite{Sanchez}. 

Apparently, the attractive potential provided by gravity will produce
essential difference of the properties of such a media. The potential
part
of the energy will produce a collapse, which
will be prevented by kinetic term. Such a competition sometimes
results in
a creation of the steady solution --- a soliton. 

It is important to mention here that the solitons in gas and fluid is
intrinsically different with e.g. single solitons considered in
particle
physics. In the case of fluid or gas only the field of density and
velocities are time independent, while the constituents of the media
are
moving. The simplest example of such a soliton is a tornado.

\section{Description of the Model}

In fluid (gas) mechanics there are two different pictures of
description.
The first, usually refereed as Eulerian, uses as the coordinates  the
space dependent fields of velocity and density. The second,
Lagrangian
description, uses the coordinates of the
particles $\vec x(\xi_i,t)$  labeled by the set of the parameters
$\xi_i$,
which could be considered as the initial positions $\vec\xi=\vec
x(\xi_i,t=0)$
 and time $t$. These initial positions  $\vec \xi$ as well, as the
coordinates $\vec x(\xi_i,t)$ belong to some domain $D \subseteq
R^3$. 
In sequel we shall consider only conservative systems, where the
paths of
different particles do not cross, therefore it is clear  that  the
functions $\vec x(\xi_i,t)$ define a diffeomorphism of  $D \subseteq
R^3$
and the inverse functions $\vec \xi(x_i,t)$ should also exist.
\begin{eqnarray}\label{1}
x_j(\xi_i,t)\Big|_{\vec \xi=\vec\xi(x_i,t)}&=x_j,\nonumber\\
\xi_j(x_i,t)\Big|_{\vec x=\vec x(\xi_i,t)}&=\xi_j.
\end{eqnarray}
The density of the particles in space at time $t$ is
\begin{equation}\label{2}
\rho(\vec x,t)=\int d^3 \xi \rho_0(\xi_i)\delta(\vec x-\vec
x(\xi_i,t)),
\end{equation}
where $\rho_0(\xi)$ is the initial density at time $t=0$.
The velocity field  $\vec v$ as a function of coordinates $\vec x$
and $t$
is:
\begin{equation}\label{3}
\vec v(x_i,t)=\dot{\vec x}(\vec\xi(x_i,t),t),
\end{equation}
where $\vec\xi(x,t)$ is the inverse function (\ref{1}). The velocity
also
could be written in the following form:
\begin{equation}\label{4}
\vec v(x_i,t)=\frac{\int d^3 \xi \rho_0(\xi_i)\dot{\vec
x}(\xi_i,t)\delta(\vec
x-\vec x(\xi_i,t))}{\int d^3 \xi \rho_0(\xi_i)\delta(\vec x-\vec
x(\xi_i,t))},
\end{equation}
or
\begin{equation}\label{5} 
\rho (x_i,t)\vec v(x_i,t)=\int d^3 \xi \rho_0(\xi_i)\dot{\vec
x}(\xi_i,t)\delta(\vec x-\vec x(\xi_i,t)).
\end{equation}
Let us calculate the time derivative of the density using its
definition
(\ref{2}) :
\begin{eqnarray}\label{6}
&\dot \rho (x_i,t)=\displaystyle\int d^3 \xi
\rho_0(\xi_i)\frac{\partial}{\partial t}
\delta(\vec x-\vec x(\xi_i,t))\nonumber\\
&=\displaystyle\int d^3 \xi \rho_0(\xi_i)\left(-\dot{\vec
x}(\xi_i,t)\right)\frac{\partial}{\partial \vec x}
\delta(\vec x-\vec x(\xi_i,t))\nonumber\\
&=-\frac{\partial}{\partial \vec x}\int d^3 \xi
\rho_0(\xi_i)\dot{\vec x}(\xi_i,t)\delta(\vec x-\vec
x(\xi_i,t))\nonumber\\
&=-\displaystyle\frac{\partial}{\partial \vec x}\rho (x_i,t)
\vec v (x_i,t)
\end{eqnarray}
In such a way we verify the continuity equation of fluid dynamics:
\begin{equation}\label{7}
\dot \rho (x_i,t)+\vec\partial\Bigl(\rho (x_i,t)\vec v
(x_i,t)\Bigr)=0.
\end{equation}.

Using the coordinates $\vec x(\xi_i,t)$ as a configurational
variables we
can consider the simplest motion of the fluid described by the
Lagrangian
\begin{equation}\label{8}
L=\int d^3 \xi \rho_0(\xi_i)\frac{m\dot{\vec x}^2 (\xi_i,t)}{2}.
\end{equation}
The equations of motion which follow from (\ref{8}) apparently are
\begin{equation}\label{9}
m\ddot{\vec x} (\xi_i,t)=0
\end{equation}
Now let us find what does this equation mean for the density and
velocity
of the fluid. For that we shall differentiate both sides of (\ref{5})
with
respect to time
\begin{eqnarray}\label{10} 
&\displaystyle\frac{\partial}{\partial t}\rho (x_i,t)\vec
v(x_i,t)=\int
d^3 \xi \rho_0(\xi_i)\ddot{\vec x}(\xi_i,t)\delta(\vec x-\vec
x(\xi_i,t))\nonumber\\
&+\int d^3 \xi
\rho_0(\xi_i)\dot{\vec
x}(\xi_i,t)\displaystyle\frac{\partial}{\partial
t}\delta(\vec x-\vec x(\xi_i,t))
\end{eqnarray}
The first term in the r.h.s. of (\ref{10}) vanishes due to the
equations
of motion(\ref{9}) and transforming the second in the same way, as we
did
in (\ref{6}) we arrive at
\begin{equation}\label{11}
\displaystyle\frac{\partial}{\partial t} \rho (x_i,t)\vec v(x_i,t)+
\displaystyle\frac{\partial}{\partial x_k}\Bigl(\rho (x_i,t)
\vec v (x_i,t) v_k(x_i,t)\Bigr)=0
\end{equation}
Let us rewrite (\ref{11}) in the following form:
\begin{eqnarray}\label{12} 
&\vec v(x_i,t)\Big[\dot \rho
(x_i,t)+\displaystyle\frac{\partial}{\partial
x_k}\Bigl(\rho (x_i,t) v_k (x_i,t)\Bigr)\Big]\nonumber\\
&+\rho(x_i,t)\Bigl[\dot{\vec
v}(x_i,t)+v_k(x_i,t)\displaystyle\frac{\partial}{\partial
x_k} \vec v (x_i,t)\Bigr]=0.
\end{eqnarray}
The first term in (\ref{12}) vanishes due to the continuity equation,
while the second gives Euler's equation in the case of the free flow:
\begin{equation}\label{13} 
\dot{\vec v}(x_i,t)+
v_k(x_i,t)\displaystyle\frac{\partial}{\partial
x_k} \vec v (x_i,t)=0
\end{equation}
In order to get the usual Euler equations for fluid or gas with the
internal pressure  we need to add to the Lagrangian
(\ref{8}) the "potential" part, as it has been shown in \cite{P.A.}.
This "potential" part describes the repulsive interaction between
particles which constitute the media (gas or fluid). As
was announced in {\bf Introduction}, we are going to introduce
another
interaction --- gravitational attraction between particles. For that
we
have to modify the Lagrangian (\ref{8}) in the following way:
\begin{equation}\label{14}
L=\int d^3 \xi \rho_0(\xi_i)\frac{m\dot{\vec x}^2
(\xi_i,t)}{2}+\frac{\gamma}{2} \int d^3 \xi d^3 \xi'
\frac{\rho_0(\xi_i)\rho_0(\xi'_i)}{|\vec x(\xi_i,t)-\vec
x(\xi'_i,t)|},
\end{equation}
where $\gamma$ denotes the gravitational constant. The equations of
motion, which follow from the Lagrangian (\ref{14}) have the form:
\begin{equation}\label{15}
m\ddot{\vec x} (\xi_i,t)+\gamma\int d^3 \xi'\rho_{0}(\xi'_i)
\frac{\vec x(\xi_i,t)-\vec x(\xi'_i,t)}{|\vec x(\xi_i,t)-\vec
x(\xi'_i,t)|^3}=0
\end{equation}
Translating equation (\ref{15}) on to the language of Euler
variables, as
has been done above we arrive at the following set, including the
continuity equation:
\begin{eqnarray}\label{16,17} 
\dot{\vec v}(x_i,t)+
v_k(x_i,t)\displaystyle\frac{\partial}{\partial x_k} \vec v (x_i,t)
&=&\frac{\gamma}{m}\displaystyle\frac{\partial}{\partial\vec x}
\int d^3 y \frac{\rho(y_i,t)}{|\vec x-\vec y|}, \\ 
\dot \rho (x_i,t)+\displaystyle\frac{\partial}{\partial \vec
x}\Bigl(\rho
(x_i,t)\vec v (x_i,t)\Bigr)&=&0.
\end{eqnarray}
Note that the r.h.s of the equation (16) in the case of ordinary gas
or
fluid is expressed through the internal pressure $p(x_i)$;
\begin{equation}\label{18}
\dot{\vec v}(x_i,t)+
v_k(x_i,t)\displaystyle\frac{\partial}{\partial x_k} \vec v
(x_i,t)=-\frac{1}{\rho (x_i)}\displaystyle\frac{\partial}{\partial
\vec
x}p(x_i)
\end{equation}
The set (16),(17) defines the evolution of initial
distribution of $\rho(x_i,t_0),
\vec v(x_i,t_0)$ of gravitating gas and,
besides it could be used to find the static configuration of this gas
for
different boundary conditions. In particular we can explore the
possibility of the existence of the static isolated configurations of
gravitating gas. Isolation here means that density $\rho(x_i)$
vanishes at
infinity. Note, that  for usual fluid, gas or plasma these kind of
solutions are forbidden due to virial arguments, known in the case of
plasma as Shafranov's theorem \cite{Shaf},\cite{Fad1}. For the gas,
describing by equation (\ref{18}) this theorem could be proven as
follows.
First, using the continuity equation let us rewrite equation
(\ref{18})
for static case in the following form:
\begin{equation}\label{19}
\displaystyle\frac{\partial}{\partial
x_k}\biggl[\rho(x_i)v_k(x_i)v_j(x_j)+\delta_{jk}p(x_i)\biggr]=0.
\end{equation}
Integrating (\ref{19}) with $x_j$ over $R^3$ we obtain:
\begin{eqnarray}\label{20}
&0=\int d^3 x x_j \displaystyle\frac{\partial}{\partial
x_k}\biggl[\rho(x_i)v_k(x_i)v_j(x_j)+\delta_{jk}p(x_i)\biggr]=\\
\nonumber
&\int d^3 x \displaystyle\frac{\partial}{\partial
x_k}\biggl(x_j\biggl[\rho(x_i)v_k(x_i)v_j(x_j)+\delta_{jk}p(x_i
\biggr]
\biggr)-\\ \nonumber
&\int d^3 x
\delta_{jk}\biggl[\rho(x_i)v_k(x_i)v_j(x_j)+\delta_{jk}p(x_i)\biggr]
\end{eqnarray}
For usual gases $p(x_i)\sim \rho^{\gamma}(x_i), \gamma >0$, therefore
the
integral
over divergence will vanish for isolated solutions for which
$\rho(x_i)\rightarrow 0$ when $|\vec x|\rightarrow \infty$ and we the
obtain the following equation:
\begin{equation}\label{21}
\int d^3 x \biggl[\rho(x_i)\vec v^2(x_j)+3 p(x_i)\biggr]=0.
\end{equation}
Apparently, this equation could be satisfied only for the case
$\rho(x_i)=0$, i.e. there is no isolated in the above formulated
sense, static solutions of the equation (\ref{18}). In order to have
a static solution of (\ref{18}) we need to change the
boundary condition $\rho(x_i)\rightarrow 0$ to the condition
$\rho(x_i)\rightarrow \rho_{as}$ , where $\rho_{as}$ is asymptotic
uniform density.

Now we shall show that the arguments of this
theorem bring no obstacles for gravitating gas. For this  we again
will rewrite the equations (16) for the static case, using continuity
equation in the following form:
\begin{equation}\label{22}
\biggl[\frac{\partial}{\partial
x_j}\biggl(\rho(x_i)v_k(x_i)v_j(x_i)\biggr)
-\frac{\gamma}{m}\rho(x_i)\frac{\partial}{\partial x_k}\int d^3 y
\frac{\rho(y_i,t)}{|\vec x-\vec y|}\biggr]=0,
\end{equation}
Integrating (\ref{22}) with $x_k$ over $R^3$ we obtain:
\begin{equation}\label{23} 
\int d^3 x x_k \biggl[\frac{\partial}{\partial
x_j}\biggl(\rho(x_i)v_k(x_i)v_j(x_i)\biggr)
-\frac{\gamma}{m}\rho(x_i)\frac{\partial}{\partial x_k}\int d^3 y
\frac{\rho(y_i,t)}{|\vec x-\vec y|}\biggr]=0.
\end{equation}
Consider the first term of the integrand in (\ref{23}). Integrating
by
parts  as above and taking into account the asymptotic conditions for
$\rho(x_i)$ we obtain:
\begin{equation}\label{24}
\int d^3 x x_k \displaystyle\frac{\partial}{\partial
x_j}\biggl(\rho(x_i)v_k(x_i)v_j(x_i)\biggr)=-\int d^3x \rho(x_i)\vec
v^2(x_i).
\end{equation}
Integration of the second term of the integrand in (\ref{23}) is 
straightforward, yielding
\begin{equation}\label{25}
-\frac{\gamma}{m}\int d^3 x x_k 
\rho(x_i)\frac{\partial}{\partial x_k}\int d^3 y
\frac{\rho(y_i,t)}{|\vec x-\vec y|}=\frac{\gamma}{2m}\int d^3 xd^3
y\frac{\rho(x_i)\rho(x_i)}{|\vec x-\vec y|} 
\end{equation}
In such a way from equation (\ref{23}) we obtain:
\begin{equation}\label{26}
\int d^3x \rho(x_i)\vec v^2(x_i)-\frac{\gamma}{2m}\int d^3 xd^3
y\frac{\rho(x_i)\rho(x_i)}{|\vec x-\vec y|}=0. 
\end{equation}
This relation apparently could be satisfied for non-trivial
configurations
of $\rho(x_i),\vec v(x_i)$. The energy functional, corresponding to
the
Lagrangian (\ref{14}) has the following form:
\begin{equation}\label{27}
E=\frac{m}{2}\int d^3x \rho(x_i,t)\vec
v^2(x_i,t)-\frac{\gamma}{2}\int
d^3 xd^3 y\frac{\rho(x_i,t)\rho(y_i,t)}{|\vec x-\vec y|}
\end{equation}
The two terms of the energy functional have clear interpretation as
kinetic $T$ and potential $U$ parts of energy and equation (\ref{23})
expresses famous "virial theorem" \cite{Landau}:
\begin{equation}\label{28}
2T=-U.
\end{equation}
Note here, that in the "virial theorem" equation (\ref{28}) holds
true for
mean values of kinetic and potential energies, while in our case of
static
solutions there is no need to average over time.

Using the relation (\ref{28}) we can easily find the total energy of
static configuration of the gravitating gas:
\begin{eqnarray}\label{29} 
E_{static}=&-\displaystyle\frac{m}{2}\int d^3x \rho(x_i)\vec
v^2(x_i)=\\
\nonumber
=&-\displaystyle\frac{\gamma}{4}\int
d^3 xd^3 y\frac{\rho(x_i,t)\rho(y_i,t)}{|\vec x-\vec y|}.
\end{eqnarray}
So, the total energy of the static solution is negative, as for the
bound
state of Kepler problem.

\section{Properties of Static Solutions}

The equations which define our static configurations of gravitating
gas have the following form:
\begin{eqnarray}\label{30}
v_k(x_i)\displaystyle\frac{\partial}{\partial x_k} \vec v (x_i)
&=&\frac{\gamma}{m}\displaystyle\frac{\partial}{\partial\vec x}
\int d^3 y \frac{\rho(y_i)}{|\vec x-\vec y|}, \nonumber\\ 
\displaystyle\frac{\partial}{\partial \vec
x}\Bigl(\rho
(x_i)\vec v (x_i)\Bigr)&=&0
\end {eqnarray}
Taking divergence of the first equation :
\begin{eqnarray}\label{31}
\vec\partial \Bigl(v_k(x_i)\displaystyle\frac{\partial}{\partial
x_k}\vec v (x_i)\Bigr)
&=&\frac{\gamma}{m}\Delta
\int d^3 y \frac{\rho(y_i)}{|\vec x-\vec y|}= \nonumber\\
&=&\frac{\gamma}{m}(-4\pi)\rho(x_i),
\end{eqnarray}
 we can write the whole set of the equations for static
configuration in a pure local form:
\begin{eqnarray}\label{32}
\vec\partial \Bigl(v_k(x_i)\displaystyle\frac{\partial}{\partial
x_k}\vec v (x_i)\Bigr)&=&-4\pi\frac{\gamma}{m}\rho(x_i),\nonumber\\
\displaystyle\frac{\partial}{\partial \vec
x}\Bigl(\rho
(x_i)\vec v (x_i)\Bigr)&=&0,\nonumber\\
\vec\partial \times
\Bigl(v_k(x_i)\displaystyle\frac{\partial}{\partial
x_k}\vec v (x_i)\Bigr)&=&0,
\end{eqnarray}
where the last equation  requires the expression
$v_k(x_i)\frac{\partial}{\partial x_k} \vec v (x_i)$ to be a
gradient.

Now we are going to derive an important inequality, which bounds the
$(-E_{static})$ of any solution of (\ref{32}) from below. For this
let us
introduce the notation for the potential $U(x_i)$:
\begin{equation}\label{33} 
U(x_i)=-\int d^3 y \frac{\rho(y_i)}{|\vec x-\vec y|}
\end{equation}
Integrating by parts we obtain the following relation:
\begin{equation}\label{34}
\int d^3 x \Bigl(\vec\partial U(x_i)\Bigr)^2=4 \pi\int
d^3 xd^3 y\frac{\rho(x_i,t)\rho(y_i,t)}{|\vec x-\vec y|}
\end{equation}
Using (\ref{34}) we can write the expression for the energy  given by
 (\ref{29}) of any static configuration, which is the solution of
(\ref{32}) in the following form:
\begin{equation}\label{35}
E_{static}=-\frac{\gamma}{16 \pi}\int d^3 x \Bigl(\vec\partial
U(x_i)\Bigr)^2
\end{equation}
Substituting into (\ref{35}) the expression for the gradient of the
potential $U(x_i)$ from the first equation (\ref{30}), we obtain the
expression for the $E_{static}$ only through the field of velocity:
\begin{equation}\label{36}
-E_{static}=\frac{m^2}{16 \pi \gamma}\int d^3 x
\Bigl(v_k(x_i)\displaystyle\frac{\partial}{\partial
x_k}\vec v (x_i)\Bigr)^2
\end{equation}
This form is most convenient for the derivation of desired
inequality. Now let us consider a function $f(x_i)$ from the Hilbert
space $W_2 ^1$, which consists of all measurable functions on $R^3$,
which have at least one derivative and square integrable on $R^3$
together with its derivatives. In particular we assume that the
density $\rho(x_i)$ belongs to $W_2 ^1$. Taking into account the
equation (\ref{31}) we have 
\begin{eqnarray}\label{37}
&|\int d^3 x \rho(x_i)f(x_i)|=|\frac{m}{4 \pi \gamma}\int d^3 x
f(x_i)\partial_k \Bigl(v_j(x_i)\partial_j v_k(x_i)\Bigr)|\nonumber \\
&=|\frac{m}{4 \pi \gamma}\int d^3 x \partial_k f(x_i)
v_j(x_i)\partial_j v_k(x_i)|\nonumber \\
&\leq\frac{m}{4 \pi \gamma}\Bigl(\int d^3 x
(\partial_k f(x_i))^2 \Bigr)^{1/2}\Bigl(\int d^3 x
(v_j(x_i)\partial_j
v_k(x_i))^2 \Bigr)^{1/2},
\end{eqnarray}
where on the last step we have used Cauchy inequality. From
(\ref{37}) we immediately obtain the inequality:
\begin{equation}\label{38}
-E_{static}\geq \pi \gamma \frac{\Bigl(\int d^3 x
\rho(x_i)f(x_i)\Bigr)^2}{\int d^3 x(\partial_k f(x_i))^2},
\end{equation}
which is valid for any $f(x_i)$ from $ W_2 ^1$ and is
saturated for $f(x_i)=U(x_i)$. Indeed, let us calculate the
derivative of the functional in the r.h.s of (\ref{38}) with respect
to $f(x_i)$:
\begin{eqnarray}\label{39} 
&\displaystyle\frac{\delta}{\delta f(x_i)}\frac{\Bigl(\int d^3 x
\rho(x_i)f(x_i)\Bigr)^2}{\int d^3 x(\partial_k
f(x_i))^2}\nonumber\\
&\displaystyle=2\frac{\Bigl(\int d^3 x
\rho(x_i)f(x_i)\Bigr)^2}{\int d^3 x(\partial_k
f(x_i))^2}\Bigl[\frac{\rho(x_i)}{(\int d^3 x
\rho(x_i)f(x_i)}+\frac{\Delta f(x_i)}{\int d^3 x(\partial_k
f(x_i))^2}\Bigr].
\end{eqnarray}
This derivative vanishes for $f(x_i)=f_{max}(x_i)$ given by 
\begin{equation}\label{40}
f_{max}(x_i)=C \Delta^{-1}\rho(x_i)=C' U(x_i),
\end{equation}
where $C$ and $C'$ are inessential constants, which do not enter
into the functional. It is easy to prove that the second variation of
this functional is negative on the $f_{max}(x_i)$, so this function
provides the absolute maximum for the functional and due to the
equations (\ref{34}-\ref{36}) its value coincides with the l.h.s. of
(\ref{38}). However, the function $f_{max}(x_i)$ does not belong to
the Hilbert space $W_2 ^1$, because the potential $U(x_i)$ has the
asymptotic behaviour $\frac{1}{|\vec x|}$ at infinity and therefore
is not square integrable. 

The integral $J$ which enters into inequality (\ref{38}) could be
written in the following form
\begin{equation}\label{41}
J=\int d^3 x \rho(x_i)f_{N}(x_i),
\end{equation}
where we denoted as $f_{N}(x_i)$ the normalized function $f(x_i)$:
\begin{equation}\label{42}
f_{N}(x_i)=\displaystyle \frac{f(x_i)}{\sqrt{{\int d^3 x
(\partial_k f(x_i))^2}}}=\frac{f(x_i)}{\|\partial_k f(x_i)\|_2}.
\end{equation}
In the Hilbert space $W_2 ^1$ the Hoelder's inequality holds true:
\begin{equation}\label{43}
|\int d^3 x f(x_i)g(x_i)|\leq
\|f\|_{p}\|g\|_{p'},\qquad\frac{1}{p}+\frac{1}{p'}=1
\end{equation}
as well, as the remarkable Ladyjenskaya's \cite{Lad} inequality:
\begin{equation}\label{44}
\|f\|_{6}\leq (48)^{1/6}\|\partial_k f\|_{2}
\end{equation}
Here we use the standard notations:
\begin{eqnarray}
\|f\|_{p}&=&\Bigl(\int d^3 x f(x_i)^{p}\Bigr)^{1/p},\nonumber\\
\|\partial_k f\|_{p}&=&\Bigl(\int d^3 x |\partial_k
f(x_i)|^{p}\Bigr)^{1/p}\nonumber
\end{eqnarray}
From these two inequalities we obtain the following bound for the
integral $J$:
\begin{equation}\label{45}
|J|\leq (48)^{1/6}\|\rho\|_{6/5}
\end{equation}
The facts mentioned above could possibly support the statement that
the ($-E_{static}$) is bounded from below by appropriate norm of the
density. Indeed, the absolute maximum of the funcional $|J|$ should
be bigger when its maximum in a restricted space like $W_2 ^1$, given
by (\ref{45}). However we can not present the rigorous proof of this
statement.

One of the other general property of a static configurations of gas
or fluid is the existence of the topological charge --- "helicity"
(or Hopf invariant), which explicit form is
\begin{equation}\label{46}
q=\int d^3 x \vec v(x_i) rot\vec v(x_i).
\end{equation}
This object is not only the integral of motion of the equations
(16) but it also is the central element of the algebra of
Poisson brackets of $(\vec v(x_i),\rho(x_i))$ \cite{P.A.}. The
existence of such an object brings additional argument for the
stability of the solitons. The role of "helicity" in the case of the
solitons in plasma was pointed out in \cite{Fad2},\cite{Fad3}.
Moreover, in \cite{Vac} it was shown that there exists a remarkable
inequality which bounds the energy of plasma solitons from below by
$q^{3/4}$. The derivation of such inequality for our case (and in
general for fluid solitons ) is highly desirable and we going to
consider this question in the future publications.

\section{On the Possible Structure of the Static Solutions}

The equations (\ref{32}) which define the static configurations of
gravitating gas are 3-dimensional nonlinear partial differential
equations and the probability to find an analytic solution is very
low. The only case where a class of solutions was found in a similar
situation is the t'Hooft-Polaykov monopole, but there the requirement
of the spherical symmetry simplified essentially the problem. In our
case we can not expect the spherically symmetric solution because the
continuity equation requires the trajectories of the particles be
closed in order to provide the static configuration for $\vec v(x_i),
\rho(x_i)$. The simplest and most symmetric configuration we could
expect for our case is the toroidal structure, where the density is
concentrated in the vicinity of the axis of the toroid, while the
field of velocity is tangential to the embedded one into the other
toroidal surfaces. It is not the first time the toroidal-shape
soliton appeared in the context of the theory of continuous media.
Since the pioneer works of Lord Kelven in XIX century to the present
time it was studied  by many scientists both mathematician and
physicists and recently the interest to the subject was again
attracted by the works of Faddeev and Niemi \cite{Fad2},\cite{Fad3}

It is clear that in the case of attractive gravitational interaction
the particles, which constitute the gas move around the region with
bigger density (like planets around the Sun) so there should exists a
collective motion in the approximation when the radius of torus tends
to infinity and we can speak about cylindrical rather when toroidal
configuration. Indeed,  let us consider this axially symmetric
tornado-like solution of (\ref{32}). For that we shall write the
Ansazt:
\begin{eqnarray}\label{47}
\vec v(x_i)&=&v(r)(-\frac{y}{r},\frac{x}{r},0),\nonumber\\
\rho(x_i)&=&\rho(r),
\end{eqnarray}
where $r=\sqrt{x^2+y^2}$. The second and the third equations
(\ref{32}) are satisfied by (\ref{47}), while the first gives
\begin{equation}\label{48}
\frac{1}{r}\partial_r v(r)^2=4\pi\frac{\gamma}{m}\rho(r).
\end {equation}
This solution shows that the density stays an arbitrary function
(what is expected for the partial differential equations) and the
velocity grows with radius up to its asymptotic value. The later is
the consequence of the approximation--- here we actually have
2-dimensional potential $log r$ instead of Coulomb $\frac{1}{r}$. 
This example is to demonstrate that the particles which
constitute the gravitating gas in their collective motion form the
localized object. 
 
The traditional way to tackle 3-dimensional gas or fluid is to
introduce for the velocity Clebsh parametrization suggested in
\cite{Clebsh} and recently discussed in\cite{Jackiw}. In
general case this parametrization has the following form:
\begin{equation}\label{49}
\vec v(x_i)=f(x_i)\vec \partial g(x_i)+\vec \partial h(x_i),
\end{equation}
where $f(x_i),g(x_i),h(x_i)$ are scalar functions.
For the toroidal solution with $z$ -- as the axis of symmetry we
shall assume that the density $\rho (x_i)$ does not depend upon the
azimuth angle $\phi$ and Clebsh parametrization takes the form:
\begin{equation}\label{50}
\vec v(x_i)=cos\alpha (r,z)\vec \partial \beta (r,z)+k\vec \partial
\phi (x_i),
\end{equation}
where $r,z,\phi$ are cylindrical coordinates,
\begin{eqnarray}
r&=&\sqrt{x^2+y^2}\nonumber\\
\phi(x_i)&=&arctang\frac{y}{x}\nonumber
\end{eqnarray}
In fluid dynamics the parametrization (\ref{50}) has an interesting
mechanical interpretation which we shall discuss elsewhere. The
"helicity" functional for this case has the following form
\begin{eqnarray}\label{51}
q=&\int k d \phi(x_i)\wedge
d cos\alpha(r,z)\wedge d\beta(r,z)\nonumber\\
=&2\pi k \int d cos\alpha(r,z)\wedge d\beta(r,z)
\end{eqnarray}
The function $\beta(r,z)$ in (\ref{50}) should has a singularity
at the point $(r_c,0)$ in the semiplane $(r,z)$  where $r_c$ is the
radius of axis line of the torus. When the point $(r,z)$ goes around
$(r_c,0)$, the function $\beta(r,z)$ increases on
$2\pi$\footnote{Compare this with the
similar parametrization of the dynamical variables in \cite{Fad3} for
the case of plasma.}. The gradient of such function will be the
vector tangential to the toroidal surfaces. The addition of  the
gradient of azimuthal angle $\phi$ makes the velocity (\ref{50})
winding also in the azimuth direction. Such a behaviour of the field
of velocity leads to the nontrivial Hopf invariant, which
characterizes the homotopy classes $\pi_3(S^2)$. Described this way
the Clebsh parameters, together with the density $\rho(x_i)$ could be
found by numerical integration of the equations (\ref{32}). 

The construction we presented and discussed above would be incomplete
and too academic without an example of the galaxy which indeed may
demonstrate the toroidal structure. Fortunately such galaxy does
exists. It was discovered back in 1950 by Art Hoag and recently a
very good picture was obtained by Hubble telescope (see figure 1). On
this picture it is clearly seen the toroidal structure and moreover
the motion of the stars seem to be very close to what we should
expect for the configuration with nontrivial "helicity" functional.
In the gallery of the galaxies which is available on the sites
http://www.astronomy.com or http://hubblesite.ogr we can find some
other examples with the form more or less close to the torus, but the
Hoag's object is the best of all. these examples show that the model
we considered might be not so trivial and could explain more
sophisticated structures like spiral galaxy.

\begin{figure}
\begin{center}
\includegraphics[490,445]{Hoag'sObject.jpg}
\end{center}
\caption{The above photo taken by the Hubble Space
Telescope in July 2001 reveals unprecedented details of Hoag's Object
and may yield a better understanding. Hoag's Object spans about
100,000 light years and lies about 600 million light years away
toward the constellation of Serpens.}
\end{figure}

\section{Acknowledgements}

I would like to thank professors A.K.Likhoded and A.V. Rasumov for
valuable discussions. The picture of Hoag's object is available
due to the courtesy of Space Telescope Science Institute. The work
was supported in part by the Russian Science Foundation Grant
04-01-00352.

\end{document}